\documentclass[aps,nopacs,nokeys,superscriptaddress,11pt,twoside,notitlepage]{revtex4-1}

\usepackage{graphicx,epic,eepic,epsfig,amsmath,latexsym,amssymb,verbatim,color}

\usepackage{theorem}
\newtheorem{definition}{Definition}

\newtheorem{lemma}[definition]{Lemma}

\newtheorem{theorem}[definition]{Theorem}

\def\squareforqed{\hbox{\rlap{$\sqcap$}$\sqcup$}}
\def\qed{\ifmmode\squareforqed\else{\unskip\nobreak\hfil
\penalty50\hskip1em\null\nobreak\hfil\squareforqed
\parfillskip=0pt\finalhyphendemerits=0\endgraf}\fi}
\def\endenv{\ifmmode\;\else{\unskip\nobreak\hfil
\penalty50\hskip1em\null\nobreak\hfil\;
\parfillskip=0pt\finalhyphendemerits=0\endgraf}\fi}

\mathchardef\ordinarycolon\mathcode`\:
\mathcode`\:=\string"8000
\def\vcentcolon{\mathrel{\mathop\ordinarycolon}}
\begingroup \catcode`\:=\active
  \lowercase{\endgroup
  \let :\vcentcolon
  }

\newcommand{\nc}{\newcommand}
\nc{\rnc}{\renewcommand}

\nc{\beg}{\begin{equation}}
\nc{\eeq}{{\end{equation}}}
\nc{\beqa}{\begin{eqnarray}}
\nc{\eeqa}{\end{eqnarray}}
\nc{\lbar}[1]{\overline{#1}}
\nc{\bra}[1]{\langle#1|}
\nc{\ket}[1]{|#1\rangle}
\nc{\ketbra}[2]{|#1\rangle\!\langle#2|}
\nc{\braket}[2]{\langle#1|#2\rangle}
\newcommand{\braandket}[3]{\langle #1|#2|#3\rangle}

\nc{\proj}[1]{| #1\rangle\!\langle #1 |}
\nc{\avg}[1]{\langle#1\rangle}
\nc{\Rank}{\operatorname{Rank}}
\nc{\smfrac}[2]{\mbox{$\frac{#1}{#2}$}}
\nc{\tr}{\operatorname{Tr}}
\nc{\ox}{\otimes}
\nc{\dg}{\dagger}
\nc{\dn}{\downarrow}
\nc{\cA}{{\cal A}}
\nc{\cB}{{\cal B}}
\nc{\cC}{{\cal C}}
\nc{\cD}{{\cal D}}
\nc{\cE}{{\cal E}}
\nc{\cF}{{\cal F}}
\nc{\cG}{{\cal G}}
\nc{\cH}{{\cal H}}
\nc{\cI}{{\cal I}}
\nc{\cJ}{{\cal J}}
\nc{\cK}{{\cal K}}
\nc{\cL}{{\cal L}}
\nc{\cM}{{\cal M}}
\nc{\cN}{{\cal N}}
\nc{\cO}{{\cal O}}
\nc{\cP}{{\cal P}}
\nc{\cQ}{{\cal Q}}
\nc{\cR}{{\cal R}}
\nc{\cS}{{\cal S}}
\nc{\cT}{{\cal T}}
\nc{\cX}{{\cal X}}
\nc{\cZ}{{\cal Z}}
\nc{\csupp}{{\operatorname{csupp}}}
\nc{\qsupp}{{\operatorname{qsupp}}}
\nc{\var}{{\operatorname{var}}}
\nc{\rar}{\rightarrow}
\nc{\lrar}{\longrightarrow}
\nc{\polylog}{{\operatorname{polylog}}}
\nc{\1}{{\openone}}
\nc{\wt}{{\operatorname{wt}}}
\nc{\av}[1]{{\left\langle {#1} \right\rangle}}

\def\g{\gamma}

\def\e{\epsilon}

\def\ll{\lambda}

\def\ps{\psi}

\nc{\RR}{{{\mathbb R}}}
\nc{\CC}{{{\mathbb C}}}
\nc{\FF}{{{\mathbb F}}}
\nc{\NN}{{{\mathbb N}}}
\nc{\ZZ}{{{\mathbb Z}}}
\nc{\PP}{{{\mathbb P}}}
\nc{\QQ}{{{\mathbb Q}}}
\nc{\UU}{{{\mathbb U}}}
\nc{\EE}{{{\mathbb E}}}
\nc{\id}{{\operatorname{id}}}

\nc{\CHSH}{{\operatorname{CHSH}}}

\nc{\be}{\begin{equation}}
\nc{\ee}{{\end{equation}}}
\nc{\bea}{\begin{eqnarray}}
\nc{\eea}{\end{eqnarray}}
\nc{\<}{\langle}
\rnc{\>}{\rangle}
\nc{\Hom}[2]{\mbox{Hom}(\CC^{#1},\CC^{#2})}
\nc{\rU}{\mbox{U}}

\nc{\ob}[1]{#1}

\nc{\SEP}{{\text{SEP}}}
\nc{\NS}{{\text{NS}}}
\nc{\LOCC}{{\text{LOCC}}}
\nc{\PPT}{{\text{PPT}}}
\nc{\EXT}{{\text{EXT}}}
\nc{\Sym}{{\operatorname{Sym}}}

\nc{\ERLO}{{E_{\text{r,LO}}}}
\nc{\ERLOCC}{{E_{\text{r,LOCC}}}}
\nc{\ERPPT}{{E_{\text{r,PPT}}}}
\nc{\ERLOCCinfty}{{E^{\infty}_{\text{r,LOCC}}}}
\nc{\Aram}{{\operatorname{\sf A}}}

\begin{document}

\title{Quantum Adiabatic Theorem Revisited}

\author{Runyao Duan}
\email{duanrunyao@baidu.com}
\affiliation{Institute for Quantum Computing, Baidu Research, Beijing 100193, China}

\begin{abstract}
In 2004 Ambainis and Regev formulated a certain form of quantum adiabatic theorem and provided an elementary proof which is especially accessible to computer scientists. Their result is achieved by discretizing the total adiabatic evolution into a sequence of unitary transformations acting on the quantum system. Here we continue this line of study by providing another elementary and shorter proof with improved bounds. Our key finding is a succinct integral representation of the difference between the target and the actual states, which yields an accurate estimation of the approximation error. Our proof can be regarded as a ``continuous" version of the work by Ambainis and Regev. As applications, we show how to adiabatically prepare an arbitrary qubit state from an initial state. 
 \end{abstract}

\maketitle

\thispagestyle{empty}

\section{Introduction} 

The quantum adiabatic theorem is a celebrated corollary of the Schr\"{o}dinger equation. Its original form was proposed by Born and Fock in 1928 \cite{BF1928}, and was generalized by Kato in 1950 \cite{Kato1950}. Since then it has been serving as a useful tool in understanding the fundamental behaviors of quantum systems under adiabatic evolution \cite{ASY1987}. The basic idea can be intuitively described as follows. Suppose a quantum system is initiated at its ground state, i.e., the eigenstate corresponding to the lowest eigenvalue of the system Hamiltonian. We now drive the system by changing the system Hamiltonian. It is well known that the system state will evolve according to the Schr\"{o}dinger equation with a time-dependent Hamiltonian, and the exact solution is usually difficult to find. Interestingly, the system will be always kept at the ground state of the system Hamiltonian at that time, providing this process is changing sufficiently smooth and long enough i.e., is an adiabatic evolution, and there is an energy gap. The quantum adiabatic theorem provides a rigorous and quantitative characterization of this intuition.

In 2000 Farhi {\it et al.} introduced a new promising model of quantum computation by adiabatic evolution whose mathematical foundation is the quantum adiabatic theorem \cite{FGGS2000}. More specifically, the authors constructed two $n$-qubit Hamiltonians $H(0)$ and $H(1)$ such that $H(0)$ has a unique easily constructible ground state (for instance, the uniform $n$-qubit state) and $H(1)$ has a unique ground state (some $n$-bit string) which encodes the solution of some combinatorial optimization problems of interest \cite{FGGS2000}. Based on the quantum adiabatic theorem, Farhi {\it et al.} showed that an interpolating Hamiltonian $H(\frac{t}{T})=(1-\frac{t}{T})H(0)+\frac{t}{T}H(1)$ will adiabatically evolve from the ground state of $H(0)$ to that of $H(1)$ within time $T$, thus providing a potential attack to classically intractable problems. It is remarkable this new model captured the full power of quantum computing \cite{AvDKL+2004}, and received increasing interests in recent years \cite{AL2018}. In particular, one of its variants, namely quantum approximate optimization algorithm (QAOA), could be either used to attack some NP complete problems \cite{FGG2014}, or to demonstrate quantum supremacy over near-term quantum computers \cite{FH2016}. 

Although the quantum adiabatic theorem was widely applied in quantum mechanics for a long time, the validity conditions of the theorem have not been fully understood. That yielded a lot of discussions on the validity of the theorem, and indeed there are many different versions of the theorem. Furthermore, most proofs techniques are usually highly involved which are only accessible to experts with heavy mathematical analysis and physics background \cite{Reichardt2004, MS2004, TSKO2005, JRS2007, EH2012}. In 2004 Ambainis and Regev rigorously formulated a certain form of the quantum adiabatic theorem, and provided an elementary proof which has captured main ingredients and key features of the adiabatic evolution \cite{AR2004}. More precisely, the authors discretized the total evolution duration into small-time intervals and approximated the whole evolution by a sequence of unitary transformations acting on these intervals. They also remarked their results could be further improved. 

Inspired by the work of Ambainis and Regev, we continue this line of study to provide another elementary and short proof with improved bounds. We employ a proof strategy similar to Ref. \cite{AR2004}, first dealing with a special case where the eigenvalue is zero ({\bf Theorem \ref{qat-s}}), and then reducing the general case to it ({\bf Theorem \ref{qat-g}}). Unlike the previous proof, we directly derive a succinct integral representation of the exact difference between the target state and the actual state, see Eq. (\ref{elegant-formula}). Based on that, we are able to estimate the approximation error accurately by the technique of integration by parts. We demonstrate applications of our results by explicitly constructing an adiabatic evolution which can be used to adiabatically prepare an arbitrary qubit from a standard state, yielding an operational interpretation of the geodesic between these two states on the Bloch sphere. We believe the simplicity of our proof make it accessible to a wider community and hope it provides new insight in developing applications on near-term quantum computers \cite{Preskill2018}.
 
 \section{Quantum Adiabatic Theorem: special case}
Suppose $A$ is a linear operator acting on a Hilbert space and $\ket{\psi}$ is a vector in the same space. We define $\|A\|:=\max_{\braket{\psi}{\psi}=1}\|A\ket{\psi}\|$ as the operator norm of $A$, where $\|\ket{\psi}\|=\sqrt{\braket{\psi}{\psi}}$ is the length of $\ket{\psi}$. $\ket{\psi}$ is a (pure) state if it is normalized, i.e., $\|\ket{\psi}\|=1$. Suppose we are given a family of Hermitian operators $\{H(s):0\leq s\leq 1\}$ such that both the first and second derivatives $H'(s)$ and $H''(s)$ are continuous with respect to $s$. We denote $\|H'\|=\max_{0\le s\le 1} \|H'(s)\|$ and $\|H''\|=\max_{0\le s\le 1} \|H''(s)\|.$

Let $\ket{\phi(s)}$ be a normalized eigenvector associated with a non-degenerate eigenvalue $\g(s)$ of $H(s)$, i.e., $H(s)\ket{\phi(s)}=\g(s)\ket{\phi(s)}$. Without loss of generality, we assume that $\ket{\phi(s)}$ is differentiable with respect to $s$. Furthermore,  as shown in Ref. \cite{AR2004}, one can choose $\braket{\phi'(s)}{\phi(s)}=0$ for $0\leq s\leq 1$ by introducing some suitable phase factor.  We assume that any other eigenvalue of $H(s)$ is at least $\ll(s)$ away from $\g(s)$, and denote $\ll=\min_{0\le s\le 1}\ll(s)$ as the minimum spectral gap. Please note that $\ket{\phi(s)}$ can be chosen as the same $k$-th largest eigenvalue of $H(s)$ for all $0\le s\le 1$ as long as the spectral gap $\ll>0$, i.e., there is no-crossing between different eigenvalues, not necessarily limited to the ground state. We will use $\phi(s)$ to represent $\ket{\phi(s)}$ directly when it does not cause any confusion. 

For $T>0$, we construct a family of Hamiltonians $\{H({t}/{T}):0\le t\le T\}$ and associate them with a quantum system. We drive this system from $H(0)$ to $H(1)$ within time $T$. If the initial state is the eigenstate of $H(0)$ with eigenvalue $\g(0)$, the final state will be close to the eigenstate of $H(1)$ with eigenvalue $\g(1)$, providing that: 1) the change is sufficiently smooth (in terms of $\|H'\|$ and $\|H''\|$); 2) the energy (spectral) gap $\ll$ is strictly positive; and 3) the evolution time $T$ is long enough. The quantum adiabatic theorem provides a quantitative relation between these quantities. More precisely, the evolution time depends on other quantities in a polynomial manner, i.e., $T=O(poly(\e^{-1},\|H'\|,\|H''\|))$. 

We follow the proof strategy in Ref. \cite{AR2004} to deal with the special case of $\g(s)=0$ first, then reduce the general case to it. We would like to emphasize this approach has greatly simplified the whole proof. In the following we assume $s=t/T$ and $dt=Tds$ without mentioning explicitly. We use ${\ps(s)}={\ps({t}/{T})}$ to represent the actual state of the quantum system at time $t$, i,e, solution to the Schr\"{o}dinger equation with a time-dependent Hamiltonian $H(t/T)$. 
 \begin{theorem}
  \label{qat-s}
 ({\bf Quantum Adiabatic Theorem: special case}) Let $\phi(s)$ be the unit eigenvector of $H(s)$ with eigenvalue $0$ and $\braket{\phi'(s)}{\phi(s)}=0$ for $0\le s\le 1$. Suppose the initial state ${\ps(0)}$ of the system is prepared at ${\phi(0)}$. Let us drive the system via the Hamiltonian $H(t/T)$ from $t=0$ to $t=T$. Then for any $\e>0$, the final state of the system $\ps(1)$ is at most $\e$ far from $\phi(1)$ in the ordinary vector nom, i.e.,  $\|{\phi(1)}-{\ps(1)}\|\le \e$, providing that 
\begin{equation}
\label{s-bound}
 T\geq \frac{1}{\e} (\frac{2\|H'\|+\|H''\|}{\lambda^2}+\frac{4\|H'\|^2}{\lambda^3}).
\end{equation}
  \end{theorem}
  
Remarks: It is evident that the evolution time $T$ depends on $\e^{-1}$, $\ll^{-1}$,  $\|H'\|$, and $\|H''\|$ polynomially. This is a quantitative characterization of our intuition of the adiabatic evolution. We also note that the condition of ``$\g(s)=0$" can be directly relaxed to ``$\g(s)=constant$". An interesting example about this useful point can be found later. 
 
 {\bf Proof:}  Two crucial steps of this proof are: (a) to establish a succinct integral representation of $\phi(1)-\ps(1)$;  (b) to further manipulate this integral representation to create a factor of $1/T$.
 
 To achieve (a),  let us first introduce $U(s_2,s_1)$, known as Dyson operator, to denote the unitary evolution from time $t_1=s_1T$ to $t_2=s_2T$. In particular, $U(s,s)=\1$ and $U(1,s)$ is the unitary evolution from time $t=sT$ to $T$. Furthermore, $U(s_2,s_1)=U(s_2,s)U(s,s_1)$ for $s_1\le s\le s_2$ . This gives the following useful fact whose validity is to be proven shortly (hereafter we use Planck units and assume $\hbar=1$): 
\begin{equation}\label{diff-u}
dU(1,s)=iTU(1,s)H(s)ds.
\end{equation}
We are now in a position to present a key finding of this paper, say, a succinct integral representation of the difference between ${\ps(1)}$ and ${\phi(1)}$, as follows: 
\begin{equation}\label{elegant-formula}
{\phi(1)}-{\ps(1)}=\int_0^1 U(1,s){\phi'(s)}ds.
 \end{equation}
 To see this, consider the vector $U(1,s){\phi(s)}$. Noting that $U(1,1)=\1$ and ${\psi(0)}={\phi(0)}$, we have 
 $${\phi(1)}=U(1,1){\phi(1)},~~{\ps(1)}=U(1,0){\ps(0)}=U(1,0){\phi(0)}.$$
By the fundamental theorem of calculus, 
 $${\phi(1)}-{\ps(1)}=U(1,s){\phi(s)}\Big|_0^1=\int_0^1 d(U(1,s){\phi(s)})=\int_0^1(dU(1,s){\phi(s)}+U(1,s)d{\phi(s)}).$$
The second term is what we want, so we only need to show that the first term vanishes. This is the case as 
$$dU(1,s){\phi(s)}=iTU(1,s)H(s){\phi(s)}ds=0,$$
where we have used Eq. (\ref{diff-u}) and $H(s)\phi(s)=0$.

It remains to show Eq. (\ref{diff-u}). This is indeed a simple property of Dyson operator and can be derived as follows:
\begin{equation}
\begin{split}\label{du-1s}
dU(1,s) &=U(1,s+ds)-U(1,s)\\
             &=U(1,s+ds)-U(1,s+ds)U(s+ds, s)\\
             &=U(1,s+ds)(\1-U(s+ds,s)).\\
\end{split}
\end{equation}
Noticing that the unitary evolution from $s$ to $s+ds$ can be approximated by a fixed Hamiltonian $H(s)$ for time $Tds$, we have
$$U(s+ds,s)=e^{-iTds H(s)}=\1-iTdsH(s)+O((ds)^2).$$ 
Substituting this into Eq. (\ref{du-1s}), we have the desired Eq. (\ref{diff-u}). (Note that we have discarded all higher order terms of $ds$). 

To achieve (b), let $P(s)=\proj{\phi(s)}$ be the projection onto ${\phi(s)}$, and $Q(s)=\1-P(s)$ be its orthogonal complement. Let $R(s)$ be the Hermitian inverse of $H(s)$ over the support of $Q(s)$, i.e., $R(s)=R^\dagger(s)$, $H(s)R(s)=Q(s)$, and $R(s)=Q(s)R(s)$. Since $\braket{\phi'(s)}{\phi(s)}=0$, we have $Q(s){\phi'(s)}={\phi'(s)}$.  So one can insert $Q(s)=H(s)R(s)$ between $U(1,s)$ and ${\phi'(s)}$ in Eq. (\ref{elegant-formula}) without changing anything, which together with Eq. (\ref{diff-u}) produces the derivative of $U(1,s)$ and a factor of $1/T$, say
 \begin{equation}\begin{split}
\label{d-for}
{\phi(1)}-{\ps(1)} &=\frac{1}{iT}\int_0^1 (iT)U(1,s)H(s)R(s){\phi'(s)}ds\\
                          &=\frac{1}{iT}\int_0^1 (iTU(1,s)H(s)ds)R(s){\phi'(s)}\\
                          &=\frac{1}{iT}\int_0^1dU(1,s)R(s){\phi'(s)}.\\
\end{split}\end{equation} 

Applying integration by parts to Eq. (\ref{d-for}), we have
\begin{equation}\label{d-formula}
{\phi(1)}-{\ps(1)}=\frac{1}{iT} (U(1,s)R(s){\phi'(s)}\Big|_0^1-\int_0^1 U(1,s)d(R(s){\phi'(s)})).
\end{equation}
Noticing that $d(R(s){\phi'(s)})=(R'(s){\phi'(s)}+R(s){\phi''(s)})ds$ and $\|UA\|=\|A\|$ for any unitary $U$, we have
\begin{equation}\label{key-inequality}
\|{\phi(1)}-{\ps(1)}\|\le \frac{1}{T}(\|R(0){\phi'(0)}\|+\|R(1){\phi'(1)}\|+\int_0^1 (\|R'(s){\phi'(s)}\|+\|R(s){\phi''(s)}\|)ds).
 \end{equation}
The above inequality indicates that 
$$\|{\phi(1)}-{\ps(1)}\|\leq \frac{C_H}{T}=O(T^{-1})$$
for some constant $C_H$ independent of $T$. So the approximation error vanishes when $T$ tends to $+\infty$. 

The rest of the proof is relatively straightforward. To make a more precise estimation of the right hand side of Eq. (\ref{key-inequality}), we use the following inequalities:
 \begin{equation}\begin{split}
 \label{chain-norms}
 \|R(s){\phi'(s)}\| &  \le \frac{\|H'(s)\|}{\ll^2(s)},\\
  \|R'(s){\phi'(s)}\| & \le \frac{2\|H'(s)\|^2}{\ll^3(s)},\\
  \|R(s){\phi''(s)}\| & \le \frac{\|H''(s)\|}{\ll^2(s)}+\frac{2\|H'(s)\|^2}{\ll^3(s)}.
 \end{split}\end{equation} 
In order to keep the main proof readable, we leave the technical but elementary proof details of Eq. (\ref{chain-norms}) to the next section. 

Substituting Eq. (\ref{chain-norms}) into Eq. (\ref{key-inequality}), we have
$$T\|{\phi(1)}-{\ps(1)}\|\le \frac{\|H'(0)\|}{\ll^2(0)}+\frac{\|H'(1)\|}{\ll^2(1)}+\int_0^1(\frac{\|H''(s)\|}{\ll^2(s)}+\frac{4\|H'(s)\|^2}{\ll^3(s)})ds.$$
Using the facts that $\ll(s)\ge \ll$, $\|H'(s)\|\le \|H'\|$, and $\|H''(s)\|\le \|H''\|$, we have 
$$\|{\phi(1)}-{\ps(1)}\|\le \frac{1}{T}(\frac{2\|H'\|+\|H''\|}{\ll^2}+\frac{4\|H'\|^2}{\ll^3})\leq\e,$$
where the last inequality is fulfilled if $T$ is chosen according to Eq. (\ref{s-bound}).
\qed
 
 \section{Upper bounding norms of individual terms}
 The purpose of this section is to complete the proof of Eq. (\ref{chain-norms}). A useful property of operator norm that will be frequently used is that $\|AB\|\le \|A\|\|B\|$ for any linear operators (or vectors) $A$ and $B$. In general we can estimate the norm of the form $R^{(m)}(s)\phi^{(n)}(s)$, where $m$ and $n$ represent the $m$-th and the $n$-th derivatives, respectively. The three inequalities in Eq. (\ref{chain-norms}) correspond to $(m,n)=(0,1), (1,1)$, and $(0,2)$, respectively.
 
By assumption on $H(s)$, the absolute value of any eigenvalue of $R(s)$ (the Hermitian inverse of $H(s)$ over the support of $Q(s)$) is at most $\frac{1}{\ll(s)}$, so $\|R(s)\|\le \frac{1}{\ll(s)}$.  Furthermore, we will make use of the following inequality from Ref. \cite{AR2004}: 
\begin{equation}
\label{d1-vec}
 \|{\phi'(s)}\|\le \frac{\|H'(s)\|}{\ll(s)}.
 \end{equation}
A slightly simpler way to see this is the following. Differentiating $H(s){\phi(s)}=0$ gives
$$H(s){\phi'(s)}=-H'(s){\phi(s)}.$$
 Left multiplying $R(s)$ to the above equation and noticing that 
 $$R(s)H(s){\phi'(s)}=Q(s){\phi'(s)}={\phi'(s)},$$ we have
 $${\phi'(s)}=-R(s)H'(s){\phi(s)}.$$
 Then Eq. (\ref{d1-vec}) follows immediately from 
 $$\|{\phi'(s)}\|\le \|R(s)\|\|H'(s)\|.$$
 
It is clear from the above proof that 
 $$\|R(s){\phi'(s)}\|\le \|R^2(s)\|\|H'(s)\|\leq \frac{\|H'(s)\|}{\ll^2(s)},$$
 which is the first inequality of Eq. (\ref{chain-norms}).
  
Now let us consider the term of $R'(s){\phi'(s)}$. We will first derive a useful expression of $R'(s)$. Differentiating $H(s)R(s)=Q(s)=\1-P(s)$ gives 
 $$H(s)R'(s)=-P'(s)-H'(s)R(s).$$ 
 Left multiplying $R(s)$, we have
 \begin{equation} 
 \label{q-rd}
 Q(s)R'(s)=-R(s)P'(s)-R(s)H'(s)R(s).
 \end{equation}
Similarly differentiating $P(s)R(s)=0$ gives 
\begin{equation}
\label{p-rd}
P(s)R'(s)=-P'(s)R(s).
\end{equation}
Adding Eqs. (\ref{q-rd}) and  (\ref{p-rd}) together yields
\begin{equation}
\label{rd}
R'(s)=-R(s)P'(s)-P'(s)R(s)-R(s)H'(s)R(s).
\end{equation}
Combining $P'(s)=\ketbra{\phi'(s)}{\phi(s)}+\ketbra{\phi(s)}{\phi'(s)}$ and $\braket{\phi'(s)}{\phi(s)}=0$ with Eq. (\ref{rd}), we have
$$R'(s){\phi'(s)}=-\braandket{\phi'(s)}{R(s)}{\phi'(s)}{\phi(s)}-R(s)H'(s)R(s){\phi'(s)}.$$
So
$$\|R'(s){\phi'(s)}\|\le |\braandket{\phi'(s)}{R(s)}{\phi'(s)}|+\|R(s)H'(s)R(s){\phi'(s)}\|\le \frac{2\|H'(s)\|^2}{\ll^3(s)},$$
which gives the second inequality of Eq. (\ref{chain-norms}).
 
 It remains to estimate the norm of $R(s){\phi''(s)}$. Differentiating twice $H(s){\phi(s)}=0$, we have
 $$H''(s){\phi(s)}+2H'(s){\phi'(s)}+H(s){\phi''(s)}=0.$$
 Left multiplying $R^2(s)$ and noticing that $R^2(s)H(s)=R(s)Q(s)=R(s)$, we have
 $$R(s){\phi''(s)}=-R^2(s)H''(s){\phi(s)}-2R^2(s)H'(s){\phi'(s)}.$$
 So $$\|R(s){\phi''(s)}\|\le \frac{\|H''(s)\|}{\ll^2(s)}+\frac{2\|H'(s)\|^2}{\ll^3(s)},$$
which is the third inequality of Eq. (\ref{chain-norms}).
\qed
Remarks: The above inequality cannot be obtained directly from Ref. \cite{AR2004}. Our proof is relatively simpler as it avoids estimating $\|{\phi''(s)}\|$. Furthermore, the result is also stronger. In fact, by applying the second claim of Lemma 3.2 in Ref. \cite{AR2004} we can only have the following weaker result: 
$$\|R(s){\phi''(s)}\|\le \|R(s)\|\|\phi''(s)\|\le \frac{\|H''(s)\|}{\ll^2(s)}+\frac{3\|H'(s)\|^2}{\ll^3(s)}.$$

 \section{reduction of the general case}
We have completed the proof of the quantum adiabatic theorem for the case of $\g(s)=0$. As shown in Ref. \cite{AR2004}, the general case of $\g(s)\neq 0$ can be reduced to this special case. The basic idea is to construct a new family of Hermitian operators $\{\widehat{H}(s):0\leq s\leq 1\}$ such that $\widehat{H}(s)=H(s)-\g(s)\1$. Then $\widehat{H}(s)$ satisfies the assumptions of Theorem \ref{qat-s}, and the eigenstate of $\widehat{H}(s)$ with eigenvalue $0$ is the same as the eigenstate of $H(s)$ with eigenvalue $\g(s)$. Furthermore, the unitary evolution induced by $\widehat{H}(s)$ is essentially equivalent to that by $H(s)$ up to a phase factor (see the Appendix for a rigorous argument for this). So we only need to provide norm relations between the derivatives of $\widehat{H}(s)$ and $H(s)$. This is quite straightforward since 
$$\|\widehat{H}'\|\le \|H'\|+\max_{0\le s\le 1} |\g'(s)|,~~\|\widehat{H}''\|\le \|H''\|+\max_{0\le s\le 1} |\g''(s)|.$$
The problem is reduced to estimate $|\g'(s)|$ and $|\g''(s)|$, which were given in Lemma 4.1 in Ref. \cite{AR2004} (with absolute value signs absent). However we found the proof of Lemma 4.1 unnecessarily involves the Taylor expansion of $\g(s)$ at $0$ and has not shown the result for general $s$. For completeness we provide a direct proof below. 

By definition, we have $H(s)\phi(s)=\g(s)\phi(s)$, so 
$\g(s)=\braandket{\phi(s)}{H(s)}{\phi(s)}$. Taking the first derivative we have
\begin{equation}\begin{split}
  \label{d1-g}
  \g'(s)  &= \braandket{\phi'(s)}{H(s)}{\phi(s)}+\braandket{\phi(s)}{H'(s)}{\phi(s)}+\braandket{\phi(s)}{H(s)}{\phi'(s)}\\
          &=\braandket{\phi(s)}{H'(s)}{\phi(s)}+\g(s)(\braket{\phi'(s)}{\phi(s)}+\braket{\phi(s)}{\phi'(s)})\\
          &=\braandket{\phi(s)}{H'(s)}{\phi(s)}+(\braket{\phi(s)}{\phi(s)})'\\
         &=\braandket{\phi(s)}{H'(s)}{\phi(s)}, \\
\end{split}
\end{equation}
where in the last equality we have used the fact that $\braket{\phi(s)}{\phi(s)}=1$ for any $0\leq s\leq 1$. (The above equation is also known as the Hellmann-Feynman relation in quantum physics literatures). So 
$$|\g'(s)|=|\braandket{\phi(s)}{H'(s)}{\phi(s)}|\le \|H'(s)\|,$$ 
and
$$\|\widehat{H}'(s)\|\le \|H'(s)\|+|\g'(s)|\le 2\|H'(s)\|~\mbox{and}~\|\widehat{H}'\|\le 2\|H'\|.$$

To derive a similar estimation of $|\g''(s)|$, we further differentiate $\g'(s)=\braandket{\phi(s)}{H'(s)}{\phi(s)}$ and obtain
$$\g''(s)=\braandket{\phi'(s)}{H'(s)}{\phi(s)}+\braandket{\phi(s)}{H''(s)}{\phi(s)}+\braandket{\phi(s)}{H'(s)}{\phi'(s)}.$$
The tricky point is that now we need to use the estimation of $\|\phi'(s)\|$ but with $\widehat{H}(s)$ instead (as $\phi(s)$ is the eigenstate of $\widehat{H}(s)$ with eigenvalue $0$). It is clear that
$$\|\phi'(s)\|\le \frac{\|\widehat{H}'(s)\|}{\ll(s)}\le \frac{2\|{H}'(s)\|}{\ll(s)}.$$
So $$|\g''(s)|\le |\braandket{\phi(s)}{H''(s)}{\phi(s)}|+2|\braandket{\phi(s)}{H'(s)}{\phi'(s)}|\le  \|H''(s)\|+\frac{4\|H'(s)\|^2}{\ll(s)}.$$
Finally we have
$$\|\widehat{H}''\|\leq \|H''\|+\max_{0\le s\le 1}|\g''(s)|\le 2\|H''\|+\frac{4\|H'\|^2}{\ll}.$$
 
Substituting the above two upper bounds into Theorem \ref{qat-s}, we have the following
 \begin{theorem}
  \label{qat-g}
 ({\bf Quantum Adiabatic Theorem: general case}) Let $\phi(s)$ be the unit eigenvector of $H(s)$ with eigenvalue $\g(s)$ and $\braket{\phi'(s)}{\phi(s)}=0$ for $0\le s\le 1$, where $\g(s)$ is twice-differentiable according to $s$. Other assumptions same as before. Then $\|{\phi(1)}-{\ps(1)}\|\le \e$, providing 
\begin{equation}\label{g-bound}
T\geq \frac{1}{\e} (\frac{4\|H'\|+2\|H''\|}{\lambda^2}+\frac{20\|H'\|^2}{\lambda^3}).
\end{equation}
  \end{theorem}

It is worth noting that when $\g(s)$ is constant, both $\g'(s)$ and $\g''(s)$ vanish. So the tighter bound in Eq. (\ref{s-bound}) of Theorem \ref{qat-s} still hold. such an example will be presented in the next section.

\section{How to adiabatically prepare a qubit state?}
Quantum adiabatic theorem provides a useful approach of preparing a target state (usually unknown or difficult to prepare) from a given initial state. Here we consider a simple case of qubit state preparation. More specifically, we will solve the following problem: 

{\bf Qubit state preparation:} Design a quantum adiabatic evolution to prepare a qubit state $\ket{\ps}$ from the initial state $\ket{0}$ up to error $\e$.

It is well known that any qubit pure state can be equivalently written into the following canonical form: 
$$\ket{\ps(\theta,\alpha)}=\cos\frac{\theta}{2}\ket{0}+e^{i\alpha}\sin\frac{\theta}{2}\ket{1},$$ 
where $0\le \theta\le \pi$ and $0\le \alpha<2\pi$. Such a state can be uniquely represented as a three-dimensional unit vector on the Bloch sphere 
$$\vec{n}(\theta,\alpha)=(n_x,n_y,n_z)=(\sin\theta \cos\alpha,\sin\theta\sin\alpha, \cos\theta).$$ 
Furthermore, $\ket{\ps(\theta,\alpha)}$ is the eigenstate of the Hermitian unitary operator $\vec{n}\cdot \vec{\sigma}$ with eigenvalue $1$, where
$$\vec{n}\cdot \vec{\sigma}=n_x\sigma_x+n_y\sigma_y+n_z\sigma_z,$$
and $\sigma_x=\ketbra{0}{1}+\ketbra{1}{0},\sigma_y=-i\ketbra{0}{1}+i\ketbra{1}{0},\sigma_z=\ketbra{0}{0}-\ketbra{1}{1}$ are Pauli matrices. 

We will give two different constructions of $H(s)$ by choosing different paths. Noticing that $\ket{\ps(\theta,\alpha)}$ and $\ket{0}$ are eigenstates of $\vec{n}\cdot \vec{\sigma}$ and $\sigma_z$ with eigenvalue $1$ respectively, we can construct the following $H(s)$ by taking the convex combination of them:
$$H(s)=(1-s)\sigma_z+s\vec{n}\cdot \vec{\sigma},~0\le s\leq 1.$$
Then $H(0)=\sigma_z$ and $H(1)=\vec{n}\cdot \vec{\sigma}$. Furthermore, one can calculate the eigenvalues of $H(s)$ as
$$\g_{\pm}(s)=\pm \sqrt{(sn_x)^2+(sn_y)^2+(1-s+sn_z)^2}=\pm\sqrt{1-2(1-s)s(1-\cos\theta)}.$$
So the spectral gap $$\ll(s)=\g_{+}(s)-\g_{-}(s)=2\sqrt{1-2(1-s)s(1-\cos\theta)},$$
and
$$\ll=\min_{s}\ll(s)=2\cos\frac{\theta}{2},$$
where the minimization is achieved for $s=1/2$.
We can easily verify that 
$$H'(s)=H(1)-H(0)=\vec{n}\cdot \vec{\sigma}-\sigma_z,$$ so 
$$\|H'\|=\|H'(s)\|=\sqrt{n_x^2+n_y^2+(n_z-1)^2}=2\sin\frac{\theta}{2},~\mbox{and}~H''(s)=0.$$
Applying the general bound in Eq. (\ref{g-bound}), we have the desired state within the error threshold $\e$ if 
$$T\ge \frac{2{\rm tg}\frac{\theta}{2}+10{\rm tg}^2\frac{\theta}{2}}{\e\cos\frac{\theta}{2}}.$$

An obvious advantage of the above construction is its simplicity in constructing $H(0)$ and $H(1)$ via a line segment. Unfortunately, it works only if $\theta\neq \pi$, i.e., the target state is not $\ket{1}$ (corresponding to the point $(0,0,-1)$ on the Bloch sphere). Indeed when $\theta=\pi$ we have $H(0.5)=0$ and the spectral gap is $0$, which makes all analysis invalid. 

To fix this issue, we will use a different path to connect $H(0)$ and $H(1)$. Noticing that $\ket{0}$ corresponds to the point $(0,0,1)=\vec{n}(0,\alpha)$ on the Bloch sphere, we can use the geodesic between them on the Bloch sphere, i,.e., the shortest arc of a large circle passing both of them, to serve as a path. That is, to choose 
$$H(s)=\vec{n}(s\theta,\alpha)\cdot \vec{\sigma}=\sin(s\theta) (\cos\alpha\sigma_x+\sin\alpha\sigma_y)+\cos(s\theta)\sigma_z, 0\le s\le 1.$$
One can readily verify that $H(0)=\sigma_z$ and $H(1)=\vec{n}\cdot \vec{\sigma}$. An important observation is that $H(s)$ always has two eigenvalues $\pm 1$ for $0\le s\leq 1$. So $\g(s)=-1$ is constant  and $\lambda=\lambda(s)=2$. Furthermore, by a direct calculation we have $\|H\|=\|H(s)\|=\theta$, and $\|H''\|=\|H''(s)\|=\theta^2$. By Theorem 1, when the evolution time $$T\ge \frac{1}{\epsilon}(\frac{1}{2}\theta+\frac{3}{4}\theta^2)=\frac{2\theta+3\theta^2}{4\e},$$
we can reach a state $\e$-close to the target state. 

The above result provides a new operational interpretation of $\theta$ as the essential cost of preparing a quantum pure state $\ket{\ps}$ from the standard state $\ket{0}$. Geometrically $\theta$ is the length of the geodesic connecting $(0,0,1)$ and $\vec{n}$ on the Bloch sphere. The above discussion can be applied directly to a general scenario of preparing a qubit pure state $\ket{\ps}$ from another initial state $\ket{\phi}$, simply replacing $\theta$ by $2\cos^{-1} |\braket{\ps}{\phi}|$. The connection between geometry and quantum computation has been observed in a previous work \cite{NDGD2006}, where the authors showed that the cost of implementing an optimal quantum circuit can be equivalently formulated as the length of the shortest path connecting two points in a certain curved geometry.

We notice that the special case of preparing $\ket{1}$, the ground state of $-\sigma_z$, from $\ket{+}=1/\sqrt{2}(\ket{0}+\ket{1})$, the ground state of $-\sigma_x$, was studied earlier by Farhi et al in their original work on quantum adiabatic computation \cite{FGGS2000}. 

\section{Discussions}

Let us briefly compare our results with some relevant previous works. Our proof can be regarded as a continuous version of Ref. \cite{AR2004}. That makes our proof is not as elementary as that in Ref. \cite{AR2004} but can give a precise estimation of the approximation error as shown in Eq. (\ref{elegant-formula}). To achieve an error threshold of $\e$, the evolution time obtained in Ref. \cite{AR2004} is 
$$T=\frac{10^5}{\e^{2}\ll^{3}}\max(\|H'\|\cdot \|H''\|,\frac{{\|H'\|}^3}{\ll}).$$ 
In contrast, the evolution time estimated by our method is given in Eq. (\ref{g-bound}), or can be simply relaxed to 
$$T=\frac{60}{\e\ll^{2}} \max(\|H'\|, \|H''\|, \frac{\|H'\|^2}{\ll}).$$
Clearly, the dependences on the error threshold $\e$, the energy (spectral) gap $\ll$, and the norm of the first derivative of Hamiltonian $\|H\|'$ have been reduced from $\e^{-2}$, $\ll^{-3}$, $\|H'\|^3$ to $\e^{-1}$, $\ll^{-2}$, $\|H'\|^2$, respectively. Furthermore, the constant in the bound has been considerably decreased.

Comparing to other proofs in existing literatures (for instance, Refs. \cite{ASY1987, Reichardt2004, JRS2007, EH2012} ), we believe that the proof presented here is much more accessible as it only uses the fundamental theorem of calculus and the technique of integration by parts. Here we want to emphasize two points which make our proof special. Firstly, in most previous proofs the so-called adiabatic evolution, i.e., a differential equation characterizing the evolution of the eigenstate $\phi(s)$, was introduced to compare with the standard evolution given by the Schr\"{o}dinger equation for $\ps(s)$. We have successfully avoided introducing that by a clever use of the evolution unitary operator $U(1, s)$ and its differential formula Eq. (\ref{diff-u}). Secondly, many previous proofs frequently employed complex analysis to introduce the so-called resolvent operator, which makes those arguments hardly accessible for researchers unfamiliar with complex analysis. Following a a successful strategy originally introduced in Ref. \cite{AR2004}, we have completely removed the use of complex analysis by first considering the special case of eigenvalue $0$, and then introducing the inverse operator $R(s)$ for $H(s)$ in the support.

Finally, it would be interesting to know whether the dependences of $\ll$ could be further reduced to $\ll^{-2}$, which is widely regarded as the optimal bound one could reach in adiabatic evolution and has been discussed in literatures (see \cite{AL2018} and references therein). Unfortunately existing approaches seem quite involved. A straightforward approach would be highly desirable.

\section*{Appendix}

Here we provide a rigorous proof of the following simple but non-trivial result which is required in the reduction of the general case to the special case.  

\begin{lemma}
Let $\widehat{H}(t)$ and $H(t)$ be two time-dependent Hamiltonians which are different up to a scalar function, say $\widehat{H}(t)=H(t)-f(t)\1$ for $t\ge 0$. Then the induced unitary evolutions $\widehat{U}(t)$ and $U(t)$ are equivalent up to a phase factor, say, $\widehat{U}(t)=e^{-ig(t)}U(t)$, where $g(t)=-\int_{0}^{t} f(s)ds$. 
\end{lemma}

{\bf Proof:} First of all, one can show that $U(t)$ satisfies the following differential equation
\begin{equation}\label{sch-u}
U'(t)=-iH(t)U(t),~U(0)=\1.
\end{equation}
This is just a rewriting of the Schr\"{o}dinger equation. Indeed, for arbitrary initial state $\ket{\ps(0)}$, the state at the time $t$ is given by 
$\ket{\ps(t)}=U(t)\ket{\ps(0)}$.
Differentiating both sides and noticing that $\ket{\ps(t)}'=-iH(t)\ket{\ps(t)}$ by the Schr\"{o}dinger equation, we have
$$U'(t)\ket{\ps(0)}=-iH(t)U(t)\ket{\ps(0)}.$$
This holds for any initial state $\ket{\ps(0)}$, thus Eq. (\ref{sch-u}) follows. 

Similarly we have 
\begin{equation}\label{sch-hatu}
\widehat{U}'(t)=-i\widehat{H}(t)\widehat{U}(t), \widehat{U}(0)=\1.
\end{equation}
Now we will choose $g(t)$ such that $\widehat{U}(t)=e^{-ig(t)}U(t)$. Clearly we need to choose $g(0)=0$ due to $\widehat{U}(0)=U(0)=\1$. To determine $g(t)$ for general $t>0$, we have
$$\widehat{U}'(t)=-ig'(t)e^{-ig(t)}U(t)+e^{-ig(t)}U'(t).$$

Substituting both Eqs. (\ref{sch-u}) and (\ref{sch-hatu}) and $\widehat{U}(t)=e^{-ig(t)}U(t)$ into the above equation and by some simple algebraic manipulations,  we have 
$$g'(t)=-f(t), g(0)=0.$$
\qed

\end{document}